# Alternative Theory of Neutrino Oscillations

Viliam Pažma[1] and Július Vanko[2]

[1]Institute of Physics, Comenius University, Bratislava, Slovakia
[2]Department of Nuclear Physics, Comenius University, Bratislava, Slovakia

vanko@fmph.uniba.sk

The formulation of the alternative theory of neutrino oscillations is presented. Also the application of that theory to a system of neutrinos produced by a source is formulated and some basic formulae are derived.

1. Introduction

In the literature on neutrino oscillations one can, from time to time, find papers dealing with problems which do not seem to be satisfactorily solved (see e.g. [1 – 5]). Some of those problems can be connected with unsatisfactorily understanding the theory of neutrino oscillations but anothers of them can represent serious objections. The existence of those (conceptual) problems indicates we need new looks at those problems or new ideas. We conjecture that, e.g., the formulation of an alternative theory can offer interesting new looks on old questions even incite their better understanding. (Naturally, such alternative theory has to exibit the acceptability to a great extent.) On account of that we tried to formulate the alternative theory of neutrino oscillations.

This paper is organized as follows. The Sec.2 cotains the formulation of the proposed theory of neutrino oscillations. The application of that theory to a system in which neutrinos are produced by a source is presented in the Secs. 3 and 4. The Sec. 5 contains several remarks on connecting problems.

2. Alternative theory of neutrino oscillations

The triplet of free neutrinos $\nu_1 = \nu_e$, $\nu_2 = \nu_\mu$, $\nu_3 = \nu_\tau$, will be described by the hamiltonian

$$H_0 = \vec{\alpha}.\vec{p} + \beta M_d ,$$
(1)

where $M_d = diag(m_1, m_2, m_3)$, $m_i$'s are masses of $\nu_i$'s and the standard physical meaning of other symbols is assumed.

The transitions $\nu_i \leftrightarrow \nu_j$ will appear in the theory if instead of $H_0$ we shall consider

$$H = H_0 + \beta M\prime = \vec{\alpha}.\vec{p} + \beta M,$$

(2)

as the generator of the time-development of neutrino states. In the last equation the matrix $M\prime$ is governed by the conditions $M\prime_{ii} = 0$, $(i = 1, 2, 3)$ and $(M\prime_{ij})^* = M\prime_{ji}$. (Because not all eigenvalues of $M$ have to be positive then $H$ has not to describe the triplet of free Dirac particles). Let us now test what result follows from these assumptions for the amplitude $A(\nu_1 \to \nu_2 ; t)$ of the transition $\nu_1 \to \nu_2$ in time $t$. If we in (2) consider the term $\beta M\prime$ as the perturbation and the initial state in $t = 0$ is the free neutrino $\nu_1 = \nu_e$ (having the momentum $\vec{p}$), then within the first order of the perturbative theory we get (the standard representation of Dirac matrices is used)

$$iA(\nu_1 \to \nu_2 ; t) = M\prime_{21} \frac{\bar{u}_2 u_1}{\sqrt{\varepsilon_1 \varepsilon_2}} e^{i \frac{\Delta\varepsilon}{2} t} \frac{\sin \frac{\Delta\varepsilon}{2} t}{\varepsilon_2 - \varepsilon_1},$$

where $\varepsilon_i = \sqrt{\vec{p}^2 + m_i^2}$, $\Delta\varepsilon = \varepsilon_2 - \varepsilon_1$, $u_i = \begin{pmatrix} \sqrt{\varepsilon_i + m_i} \; w_- \\ -\sqrt{\varepsilon_i - m_i} \; w_- \end{pmatrix}$,

(we choose $\vec{p} = (0, 0, p \rangle 0)$ and $\sigma_3 w_- = -w_-$).

Now we get for the probability $P(\nu_1 \to \nu_2 ; t)$ of the considered transition the expression

$$P(\nu_1 \to \nu_2 ; t) = \frac{2|M\prime_{21}|^2 \left(\varepsilon_1 \varepsilon_2 + m_1 m_2 - \vec{p}^2\right)}{\varepsilon_1 \varepsilon_2 (\varepsilon_2 - \varepsilon_1)^2} \sin^2 \frac{\Delta\varepsilon t}{2}.$$

If we confine ourselves to the region $\vec{p}^2 \gg m_i^2$, then

$$P(\nu_1 \to \nu_2 ; t) = \frac{4|M\prime_{21}|^2}{(m_2 - m_1)^2} \sin \frac{\Delta m^2 t}{4p},$$

(3)

where $\Delta m^2 = m_2^2 - m_1^2$.

The standard theory of neutrino oscillations for $P$ offers this expression

$$P(v_1 \to v_2 ; t) = \frac{4|M'_{21}|^2}{(m_2 - m_1)^2 + 4|M'_{21}|^2} \sin \frac{\Delta m^2 t}{4p},$$

if only two neutrinos $v_1, v_2$ are taken into account.

At this step we can conclude that the standard theory and the presented one do not offer fundamentally different results in the region $\vec{p}^2 \gg m_i^2$ and in the framework of the first order of the perturbative theory at least.

3. Oscillations of neutrinos produced by a source.

In the standard theory and the presented one the time-development of states is determined by the equation

$$i\partial_t \Psi - H\Psi = 0 .$$

(4)

However, if we should want to derive some forecasts relating to, for instance, solar neutrinos by means of (4) we immediatly encounter a difficulty. Namely, the Sun is the source of neutrinos thus their number does not conserve. This simply means that the density $\rho$ of neutrinos and the density $\vec{j}$ of their flux cannot be governed by the equation

$$\partial_t \rho + \text{div } \vec{j} = 0 .$$

However, the last equation follows from (4) if $H^+ = H$. Hence, if we deal with a source of neutrinos then the continuity equation has to be of the form

$$\partial_t \rho + \text{div } \vec{j} = n(\vec{x}, t) ,$$

(5)

where $n(\vec{x}, t)$ is the density of neutrinos produced per unit of time.

The equation (5) can be obtained from (4) if we reject the requirement $H^+ = H$. However, if we admit $H^+ \neq H$ then we encounter the difficulty relating to the unitarity condition. On account of that we propose the equation of the type

$$i\partial_t \Psi - H\Psi = i\varphi_0 , \quad (H^+ = H) ,$$

(6)

as the equation describing particles produced by a source. In the last equation the function $\varphi_0(\vec{x}, t)$ is determined by the properties of a source.

It seems to us natural to assume that any source can be represented as a set of point sources distributed in some region with the density $\rho_s(\vec{x}, t)$. A point source situated at $\vec{x}$ in time $t$ produces $N(\vec{x}, t)$ neutrinos per unit of time. The spectrum of neutrinos produced by each point of the source is described by the amplitude $C(\vec{p},\vec{x},t)$. On the basis of this idea we expect that $\varphi_0$ has to be expressed by means of $\rho_s$, $N$, $C$. It will be done below.

Let us now return to the equation (6). Ignoring neutrinos which are not produced by a source in question then the solution to (6) can be written in the form

$$\Psi(\vec{x},t) = \int_{-\infty}^{t} dt' \, e^{-iH(t-t')} \varphi_0(\vec{x},t') .$$

(7)

If $\varphi_0$ is independent on time (i.e. $\rho_s$, $N$, $C$ are independent on time and hence we deal with a stationary source) then from (7) we obtain

$$\Psi(\vec{x},t) = -iH^{-1} \varphi_0(\vec{x}) \equiv \Psi(\vec{x}) .$$

(8)

So in the surroundings of a stationary source the quantities as e.g. $\rho$, $\vec{j}$, .... are independent on time as necessary. (Naturally, a sufficiently long time is need to set such physical situation on).

As to $\varphi_0$ we propose the following relation among $\varphi_0$, $\rho_s$, $N$, $C$ (we confine ourselves to a stationary source composed of the identical point sources producing neutrinos $\nu_1$ type only)

$$\varphi_0(\vec{x}) \sim \int d^3\vec{x}_0 \, \rho_s(\vec{x}_0) \, N(\vec{x_0}) \int \frac{d^3\vec{p}}{(2\pi)^{\frac{3}{2}}} C(\vec{p}) \, \varepsilon_1 \frac{U_1}{\sqrt{2\varepsilon_1}}$$

$e^{i\vec{p}\cdot(\vec{x}-\vec{x}_0)}$,   (9)

where

$$U_1 = \begin{pmatrix} u_1 \\ 0 \\ 0 \end{pmatrix}, \quad U_2 = \begin{pmatrix} 0 \\ u_2 \\ 0 \end{pmatrix}, \quad U_3 = \begin{pmatrix} 0 \\ 0 \\ u_3 \end{pmatrix},$$

and bispinors $u_i$'s are governed by the equations

$$u_i \, \varepsilon_i = (\vec{\alpha}\cdot\vec{p} + m_i\beta) \, u_i$$

and correspond to the negative helicity. The missing constant of proportionality in (9) will not play role in our next considerations and thus we shall ignore it.

Let us now imagine a single stationary point source (placed at $\vec{x}_0$) which produces neutrinos with momentum $\vec{p}$ only. Now

$$\varphi_0(\vec{x}) \sim \varepsilon_1 \frac{U_1}{\sqrt{2\varepsilon_1}} e^{i\vec{p}\cdot(\vec{x}-\vec{x}_0)} .$$

Ignoring the oscillations ($\beta M' = 0$) then we get

$$\Psi(\vec{x}) \sim -iH_0^{-1}\varphi_0 = \frac{U_1}{\sqrt{2\varepsilon_1}} e^{i\vec{p}\cdot(\vec{x}-\vec{x}_0)} .$$

This result seems to be fully acceptable. (To obtain this result we were forced to put the factor $\varepsilon_1$ into (9)).

4. Probabilities of oscillations

If a source produces the neutrinos $\nu_1$ only and the time-development is governed by (6) then at the point $\vec{x}$ we can register $\nu_k$ with certain probability. In the next we shall derive expressions for those probabilities. Denoting

$$\int d^3\vec{x}_0 \, \rho_s(\vec{x}_0) \, N(\vec{x}_0) \, e^{-i\vec{p}\cdot\vec{x}_0} = F(\vec{p})$$

and

$$-iH^{-1}C(\vec{p})\,\varepsilon_1 \frac{U_1}{\sqrt{2\varepsilon_1}} e^{i\vec{p}\cdot\vec{x}} = \sum_{k=1}^{6} C_k(\vec{p}) \frac{U_k}{\sqrt{2\varepsilon_k}} e^{i\vec{p}\cdot\vec{x}} ,$$

where $U_k$ for $k = 4, 5, 6$ represent solutions corresponding to negative energies then

$$\Psi(\vec{x}) = \sum_{k=1}^{6} \Psi_k(\vec{x}) = \sum_{k=1}^{6} \int \frac{d^3\vec{p}}{(2\pi)^{\frac{3}{2}}} F(\vec{p}) \, C_k(\vec{p}) \frac{U_k}{\sqrt{2\varepsilon_k}} e^{i\vec{p}\cdot\vec{x}} .$$

(10)

Now if we put $\vec{x} = \vec{n}L$, ($\vec{n}^2 = 1$), $L \gg$ linear dimension of a source, then we shall register $\nu_k$ at $\vec{x} = \vec{n}L$ with the probability $P(\nu_1 \to \nu_k; \vec{n}L)$ which is equal to the ratio

$$P(v_1 \to v_k; \vec{n}L) = \frac{\left|\vec{j}_k^{(\vec{n})}(\vec{n}L)\right|}{\left|\vec{j}_{tot}^{(\vec{n})}(\vec{n}L)\right|},$$

(11)

where $\vec{j}_k^{(\vec{n})}(\vec{n}L)$ is the density of flux of neutrinos $v_k$ having momenta $\vec{p} = p\vec{n}$, ($p \in \langle 0, \infty \rangle$) and $\vec{j}_{tot}^{(\vec{n})}(\vec{nL})$ can be expressed as

$$\vec{j}_{tot}^{(\vec{n})} = \sum_{k=1}^{6} \vec{j}_k^{(\vec{n})}(\vec{nL}).$$

We note that the quantities $\vec{j}_e^{(\vec{n})}$ and $\vec{j}_{\mu+\tau}^{(\vec{n})}$ were measured, for instance, in SNO experiment [6, 7].

Because neutrino $v_k$ is described by

$$\Psi_k(\vec{x}) = \int \frac{d^3\vec{p}}{(2\pi)^{\frac{3}{2}}} F(\vec{p}) C_k(\vec{p}) \frac{U_k}{\sqrt{2\varepsilon_k}} e^{i\vec{p}\cdot\vec{x}} =$$

$$\int d\Omega(\vec{n}) \int_0^\infty \frac{dp\, p^2}{(2\pi)^{\frac{3}{2}}} F(p\vec{n}) C_k(p\vec{n}) \frac{U_k}{\sqrt{2\varepsilon_k}} e^{ip\vec{n}\cdot\vec{x}} =$$

$$\int d\Omega(\vec{n}) \Psi_k^{(\vec{n})}(\vec{n}.\vec{x}),$$

then

$$\vec{j}_k^{(\vec{n})}(\vec{n}L) = \Psi_k^{+(\vec{n})}(L)\, \vec{\alpha}\, \Psi_k^{(\vec{n})}(L).$$

It is evident that (11) is idependent on a constant of the proportionality which is missing in (9).

## 5. Concluding remarks

Up to now we have no results concerning any physically relevant system. Now we try, first of all, for better understanding the presented ideas and revealing their weak (or not quite clear) points.

Inciting ideas for previous considerations there were the following ones :

a) Studying solar neutrino oscillations we deal with their source. That source can be considered with the high accuracy as stationary one. However, physical quantities in the surroundings (which also does not change with time) of that source cannot depend on time as it follows from our experiences. (The presented considerations render such result).

b) Let us imagine a source which produced neutrinos in times $t_1 < t_2 < t_3 < .... < t_n$. It is natural to expect that the wave function $\Psi$ of produced neutrinos in time $t > t_n$ must be equal to

$$\Psi(\vec{x},t) = \sum_i \Psi_{t_i}(\vec{x},t) .$$

In this formal record the quantity $\Psi_{t_i}(\vec{x},t)$ represents the contribution to $\Psi$ coming from the emission of neutrinos in time $t_i$. This idea seems to be transparent and acceptable and we conjecture that (7) represents it.

For the time being we do not want compare these two theories although there are several points of contact but also many fundamental distinctions.

This work was supported by the Slovak grant agenture VEGA, Contract No.1/8315/01.